\documentclass[twocolumn,amsmath,amssymb,pra,footinbib]{revtex4-1}
\usepackage{color}
\usepackage[pdftex]{graphicx}
\usepackage{dsfont}
\usepackage{mathrsfs}
\usepackage[unicode=true,bookmarks=false, breaklinks=false,pdfborder={0 0 1},backref=false,colorlinks=true]{hyperref}

\input epsf
\begin{document}
\title {Collective phases of strongly interacting cavity photons}
\author{Ryan M. Wilson$^{1}$, Khan W. Mahmud$^{2}$, Anzi Hu$^{3}$, Alexey V. Gorshkov$^{2,4}$, Mohammad Hafezi$^{2,5,6}$, and Michael Foss-Feig$^{2,7}$}
\affiliation{$^1$Department of Physics, The United States Naval Academy, Annapolis, MD 21402, USA}
\affiliation{$^2$Joint Quantum Institute, NIST/University of Maryland, College Park, MD 20742, USA}
\affiliation{$^3$Department of Physics, American University, Washington, DC 20016, USA}
\affiliation{$^4$Joint Center for Quantum Information and Computer Science, NIST/University of Maryland, College Park, Maryland 20742, USA}
\affiliation{$^5$Department of Electrical Engineering and Institute for Research in Electronics and Applied Physics, University of Maryland, College Park, MD 20742, USA}
\affiliation{$^6$Kavli Institute of Theoretical Physics, Santa Barbara, CA 93106, USA}
\affiliation{$^7$United States Army Research Laboratory, Adelphi, MD 20783, USA}
\begin{abstract}
We study a coupled array of coherently driven photonic cavities, which maps onto a driven-dissipative XY spin-$\frac{1}{2}$ model with ferromagnetic couplings in the limit of strong optical nonlinearities.  Using a site-decoupled mean-field approximation, we identify steady state phases with canted antiferromagnetic order, in addition to limit cycle phases, where oscillatory dynamics persist indefinitely. We also identify collective bistable phases, where the system supports two steady states among spatially uniform, antiferromagnetic, and limit cycle phases. We compare these mean-field results to exact quantum trajectories simulations for finite one-dimensional arrays. The exact results exhibit short-range antiferromagnetic order for parameters that have significant overlap with the mean-field phase diagram.  In the mean-field bistable regime, the exact quantum dynamics exhibits real-time collective switching between macroscopically distinguishable states. We present a clear physical picture for this dynamics, and establish a simple relationship between the switching times and properties of the quantum Liouvillian.  
\end{abstract}
\maketitle

Despite {numerous} outstanding questions, the study of quantum many-body systems in thermal equilibrium is on relatively solid ground.  In particular, very general guiding principles help to categorize the possible equilibrium phases of matter, and predict in what situations they can occur~\cite{Fisher74,RevModPhys.70.653,sachdev2007quantum}. 
In comparison, quantum many-body systems that are far from equilibrium are less thoroughly understood, motivating a large scale effort to explore non-equilibrium dynamics experimentally, in particular using atoms, molecules, and photons~\cite{Kasprzak06,Hofferberth07,Amo08,Amo09,Deng10,Polkovnikov11}.  At the same time, it has become clear that studying non-equilibrium physics in these systems is often more natural than studying equilibrium physics; they are, in general, \emph{intrinsically} non-equilibrium.  For example, thermal equilibrium is essentially never a reasonable assumption in photonic systems, where dissipation must be countered by active pumping~\cite{RevModPhys.85.299}.  Indeed, the inadequacy of equilibrium descriptions for photonic systems has long been recognized~\cite{PhysRev.159.208}, even though close analogies to thermal systems sometimes exist~\cite{PhysRevA.2.1170, PhysRevA.87.023831,Kirton13,Hafezi15,Maghrebi16}.    

Until recently, photonic systems have been restricted to a weakly interacting regime.  With notable progress towards generating strong optical nonlinearities at the few-photon level, for example with atoms coupled to small-mode-volume optical devices~\cite{Spillane08,PhysRevLett.104.203603,Hennessy07,Nature_Photon_Phase_Gate,Hafezi12,Thompson13}, Rydberg polaritons~\cite{PhysRevLett.107.133602, Nature_Rydberg_EIT}, and circuit-QED devices~\cite{Houck12,Nissen12,Raftery14,Barends15}, this situation is rapidly changing.  The production of strongly interacting, driven and dissipative gases of photons appears to be feasible~\cite{Chang08,Chang14}, and affords exciting opportunities to explore the properties of open quantum systems in unique contexts, while studying the applicability of theoretical treatments designed with more weakly interacting systems in mind.  For example, it is not fully understood how the steady states of these systems relate to the equilibrium states of their ``closed'' counterparts, or how  conventional optical phenomena, such as bistability, manifests in the presence of strong optical nonlinearities and spatial degrees of freedom. 

We consider an array of coupled, single-mode photonic cavities described by a driven-dissipative Bose-Hubbard model~\cite{Hartmann06,Hartmann08rev,Carusotto09,Leboite13,Leboite14}, which maps onto a driven-dissipative XY spin-$\frac{1}{2}$ model in the limit of strong optical nonlinearity.  We perform a comprehensive mean-field (MF) study, and identify a variety of interesting steady states including spin density waves and limit cycles, which break the discrete translational symmetry of the system.   The spin density waves possess canted antiferromagnetic order for a range of drive strengths, despite the ferromagnetic nature of the spin couplings.  Interestingly, the exact quantum solutions exhibit short-range antiferromagnetic correlations for parameters that have notable overlap with the MF results. 
The system also supports collective bistable phases, which manifest in the exact quantum dynamics as fluctuation-induced collective switching between MF-like states.  We present a simple relationship between this dynamics and properties of the quantum Liouvillian. 

\section{Model} For a system weakly coupled to a Markovian environment, the dynamics of its density matrix $\hat{\rho}$ is governed by a master equation  $\partial_t \hat{\rho} =  {\mathscr{L}}\left[ \hat{\rho} \right] $, where ${\mathscr{L}} \left[ \hat{\rho} \right] = - i [ \hat{\mathcal{H}}  , \hat{\rho} ] + {\mathscr{D}}\left[ \hat{\rho} \right]$ is the Liouvillian, $\mathcal{\hat{H}} $ is the system Hamiltonian, and $\mathscr{D}[\hat{\rho}]$ is a dissipator in the Lindblad form~\cite{Breuertext,Carmichaeltext}.  Here, we consider an array of coherently coupled, nonlinear, single-mode photonic cavities driven by a spatially uniform laser field with frequency $\omega_\mathrm{l}$, which leak photons into the environment at a rate $\gamma$.  In the frame rotating at the driving frequency, the Hamiltonian for the coherent drive is $\hat{\mathcal{H}}_\mathrm{l} =   \Omega \sum_i (  \hat{a}_i + \hat{a}^\dagger_i ) $.  The system Hamiltonian is then $\hat{\mathcal{H}} = \hat{\mathcal{H}}_\mathrm{BH} + \hat{\mathcal{H}}_\mathrm{l}$, where
\begin{align}
\label{Hamiltonian}
\hat{\mathcal{H}}_\mathrm{BH} &= - \frac{J}{d} \sum_{ \langle i , j  \rangle } \hat{a}^\dagger_i \hat{a}_j - \mu \sum_i \hat{n}_i  +  \frac{U}{2} \sum_i \hat{n}_i \left( \hat{n}_i - 1 \right) 
\end{align}
is the Bose-Hubbard (BH) Hamiltonian~\cite{Fisher89}.  Here, $\hat{a}_i^\dagger$ ($\hat{a}_i$) creates (annihilates) a photon in cavity $i$,  $\hat{n}_i = \hat{a}^\dagger_i \hat{a}_i $ is the local number operator, and the sums run from $i=1$ to $\mathcal{N}$, the number of cavities.  The notation $\langle i , j \rangle$ implies a further summation over all cavities $j$ that are nearest-neighbors to cavity $i$, the number of which is given by the coordination number $z$.  The nonlinearity of the cavities is quantified by effective two-photon interactions of strength $U$. The first term in Eq.~(\ref{Hamiltonian}) describes the hopping of photons between cavities; $J$ is the hopping rate, and $d$ is the dimensionality of the system ($d = z/2$ for hypercubic arrays).  The laser drives the system with strength $\Omega$, and is detuned from the cavity resonance frequency $\omega_\mathrm{c}$ by $\mu =  \omega_\mathrm{l} - \omega_\mathrm{c} $.  The coupling of the system to the environment is described by the dissipator ${\mathscr{D}} [ \hat{\rho} ]=\frac{\gamma}{2} \sum_i ( 2 \hat{a}_i \hat{\rho} \hat{a}^\dagger_i - \hat{\rho} \hat{n}_i - \hat{n}_i \hat{\rho})$.

\begin{figure}[t!]
\includegraphics[width=\columnwidth]{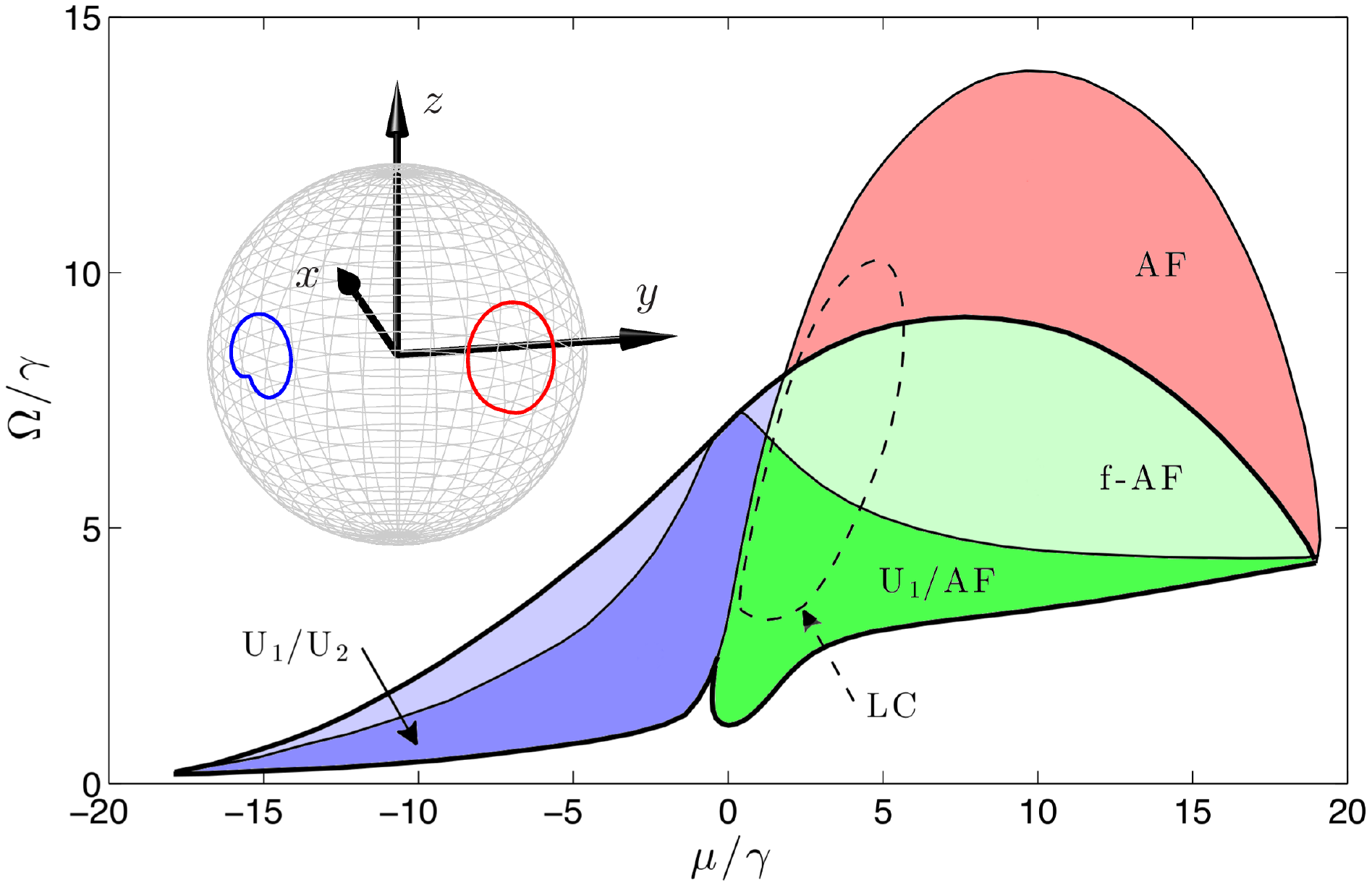}  
\caption{\label{fig1} (color online).   Mean-field phase diagram for $J/\gamma = 10$ in the hard-core ($U\rightarrow \infty$) limit.   Represented are dark ($U_1$) and bright ($U_2$)  uniform states, canted antiferromagnetic states (AF), frustrated AF states (f-AF), and limit cycles (LC).  Regions with double labeling exhibit bistability between the indicated states.   Discontinuous transitions are indicated by  thick black lines.  The inset shows a limit cycle for $\mu/\gamma = 2.5$ and $\Omega / \gamma = 6.5$ projected onto the Bloch sphere.  The red (blue) lines represent the dynamics of cavities in the A (B) sublattice.}
\end{figure}

For $\Omega=0$, the system evolves into a trivial vacuum at long times, with $\langle \hat{n}_i \rangle = 0$ for all $i$. The laser drive provides a photon  source, and can stabilize non-trivial steady states $\hat{\rho}^\mathrm{ss}$, which satisfy ${\mathscr{L}}[ \hat{\rho}^\mathrm{ss} ] = 0$.  Generally, these non-equilibrium steady states are qualitatively distinct from the equilibrium states of the closed ($\gamma=\Omega=0$) BH model described by $\hat{\mathcal{H}}_\mathrm{BH}$, which are characterized by a superfluid order parameter that spontaneously breaks the $U(1)$ symmetry associated with particle number conservation.  This $U(1)$ symmetry is explicitly broken by the coherent laser drive, and the driven-dissipative BH (DDBH) model conserves neither energy nor particle number.  Therefore superfluidity cannot emerge in the DDBH model; this is in contrast to similar models with incoherent pumps, which can support superfluid phases~\cite{Greentree06,Angelakis07,Hartmann07,Rossini07,Koch09,Diehl10,deLeeuw15}.  The DDBH model does, however, possess a spatial symmetry generated by discrete translations along the Bravais vectors of the cavity array, which is broken spontaneously if spatial structure develops in the steady state. 

 We study the steady states of the DDBH model using both a site-decoupled Gutzwiller mean-field (MF) method~\cite{Rokhsar91} and exact quantum trajectory simulations of finite systems~\cite{Dalibard92,Dum92,Plenio98,Daley14}.  In this MF approximation, the density matrix is decomposed as a matrix product $\hat{\rho} = \bigotimes \hat{\rho}_i $, where $\hat{\rho}_i$ is the local density matrix at cavity $i$.  Further, we restrict our study to the ``hard-core'' limit, where strong optical nonlinearities produce a perfect photon blockade, by taking $U \rightarrow \infty$~\cite{Carusotto09}.   In this limit, the photons can be mapped onto spins by an inverse Holstein-Primakoff transformation~\cite{Holstein40}, resulting in an effective driven-dissipative XY spin-$\frac{1}{2}$ model~\cite{Angelakis07,Joshi13} with symmetric, ferromagnetic spin couplings, described by the Hamiltonian $\hat{\mathcal{H}} = -\frac{J}{ 4 d} \sum_{\langle i , j \rangle} \left( \hat{\sigma}^x_i \hat{\sigma}^x_j + \hat{\sigma}^y_i \hat{\sigma}^y_j   \right)  + \Omega \sum_i \hat{\sigma}^x_i - \frac{\mu}{2} \sum_i \hat{\sigma}^z_i$, where  $\hat{\sigma}^{x,y,z}_i$ are the Pauli matrices.  We derive equations of motion for the spin components $\sigma^\alpha_i = \mathrm{Tr}[ \hat{\rho} \hat{\sigma}^\alpha_i ]$; these are given by Eqs.~(\ref{sspineom}) in the appendix.  For clarity, we specialize to the case of $J/ \gamma=10,$ though many of the qualitative features discussed below are valid more generally.
 
\section{Mean-field phase diagram}  By solving the MF equations for a variety of parameters, we find heuristically that all steady states are either spatially uniform, where all spins point in the same direction ($\sigma^\alpha_i = \sigma^\alpha_j$ for all $i,j$), or have antiferromagnetic (AF) spin density wave order, where neighboring spins point in different directions, but next nearest neighbors point in the same direction ($\sigma^\alpha_i \neq \sigma^\alpha_{i \pm 1}$ and $\sigma^\alpha_i  = \sigma^\alpha_{i \pm 2}$ for all $i$).  Because the neighboring spins are not antiparallel, this AF order is ``canted.''  This motivates the use of a two-sublattice \emph{ansatz}, which we solve by evolving the MF equations of motion on two sites.  We find a variety of interesting steady state phases, shown by the colored regions in Fig.~\ref{fig1}. {In the blue region, the system exhibits bistability between spatially uniform darker ($\mathrm{U}_1$) and brighter ($\mathrm{U}_2$) steady states.  In the red region, there is a unique steady state with AF order.  In the green region, the system is bistable between $\mathrm{U}_1$ and $\mathrm{AF}$ steady states.   All phase boundaries in Fig.~\ref{fig1} correspond to continuous transitions except those at the threshold of {bistability (dark lines), where the additional steady state appears discontinuously.}  
 
The bistability in this system is inherently \emph{collective}, in that it does not exist for a single cavity in the hard-core limit~\cite{Drummond80}.  We note that collective bistability exists in a variety of other driven and dissipative systems~\cite{Bonfacio78,Bowden79,Carmichael86,Sarkar87,Nagorny03,Elsasser04,Armen06,Lee12,Marcuzzi14}, and was recently observed in an gaseous ensemble of Rydberg atoms~\cite{Carr13}.  Gases of Rydberg atoms are also predicted to exhibit AF order~\cite{Lee11,Hu13,Hoening14}, though unlike the model we consider here, their interactions (due to the Rydberg blockade) are effectively antiferromagnetic in nature.  Other works studying the hard-core DDBH model also predict AF order, though they consider variants of the model that include spatially varying drive fields~\cite{Hartmann10}, two-cavity pumping~\cite{Joshi13}, and cross-Kerr terms~\cite{Jin13,Jin14}.  Our system exhibits AF order in the absence of these features, despite the ferromagnetic nature of the couplings.
 
 In the region enclosed by the dashed line in Fig.~\ref{fig1}, {the AF states acquire a limit cycle (LC) character~\cite{Nicolistext,Lee11,Qian12}, exhibiting periodic oscillatory dynamics at long times, and thus breaking the continuous time-translational symmetry of the system.
The inset in Fig.~\ref{fig1} shows an example limit cycle projected onto a Bloch sphere.   Interestingly, the limit cycle exists both as a unique steady state, and as part of a $\mathrm{U}_1 / \mathrm{LC}$ bistable pair {(green region)}.  We note that similar limit cycles were recently predicted for a driven-dissipative XYZ spin-$\frac{1}{2}$ model~\cite{Chan15}. 

To determine the validity of the two-sublattice \emph{ansatz}, we perform a linear stability analysis on the spatially uniform steady states~\cite{Leboite13,Joshi13,Leboite14} (see Sec.~\ref{sec:linstab} in the appendix).  In the light blue and light green regions in Fig.~\ref{fig1}, the $\mathrm{U}_1$ steady state is dynamically unstable to the formation of incommensurate ($k<\pi$) spin density waves, which cannot be captured within the two-sublattice \emph{ansatz}.  To understand the effects of this instability, we solve the inhomogeneous MF equations for a 1D chain, with randomly seeded initial states.  In the light blue region, the instability results in the disappearance of the $\mathrm{U}_1$ steady state.  In the light green region, we find steady states that exhibit AF order over finite size domains, which are interrupted by domains of the $\mathrm{U}_1$ state; we refer to this as frustrated AF order (f-AF).  While the AF phase is the only true periodically ordered steady state in this region, $\mathrm{U}_1$ domains remain stable if they are sufficiently small, and do not sample unstable wave vectors that often exist over a very narrow range of $k$. The limit cycles remain {mostly} AF ordered, with some low amplitude, small-$k$ features, and the $\mathrm{U}_1$ state of the $\mathrm{U}_1 / \mathrm{LC}$ pair ceases to exist in the light green LC region.

\section{Beyond mean-field}  It is important to understand how the rich physics predicted by the Gutzwiller MF theory survives in the presence of quantum and classical fluctuations, which exist in the true steady state of the DDBH model, and play particularly important roles in low dimensions.  Toward this end, we employ a quantum trajectories algorithm to study finite 1D systems~\cite{Dalibard92,Dum92,Plenio98,Daley14} (see Sec.~\ref{sec:jumps} in the appendix).  This method provides exact results for physical observables in the steady state under the ensemble averaging of trajectories.

\begin{figure}[t]
\hspace*{-2pt} \includegraphics[width=\columnwidth]{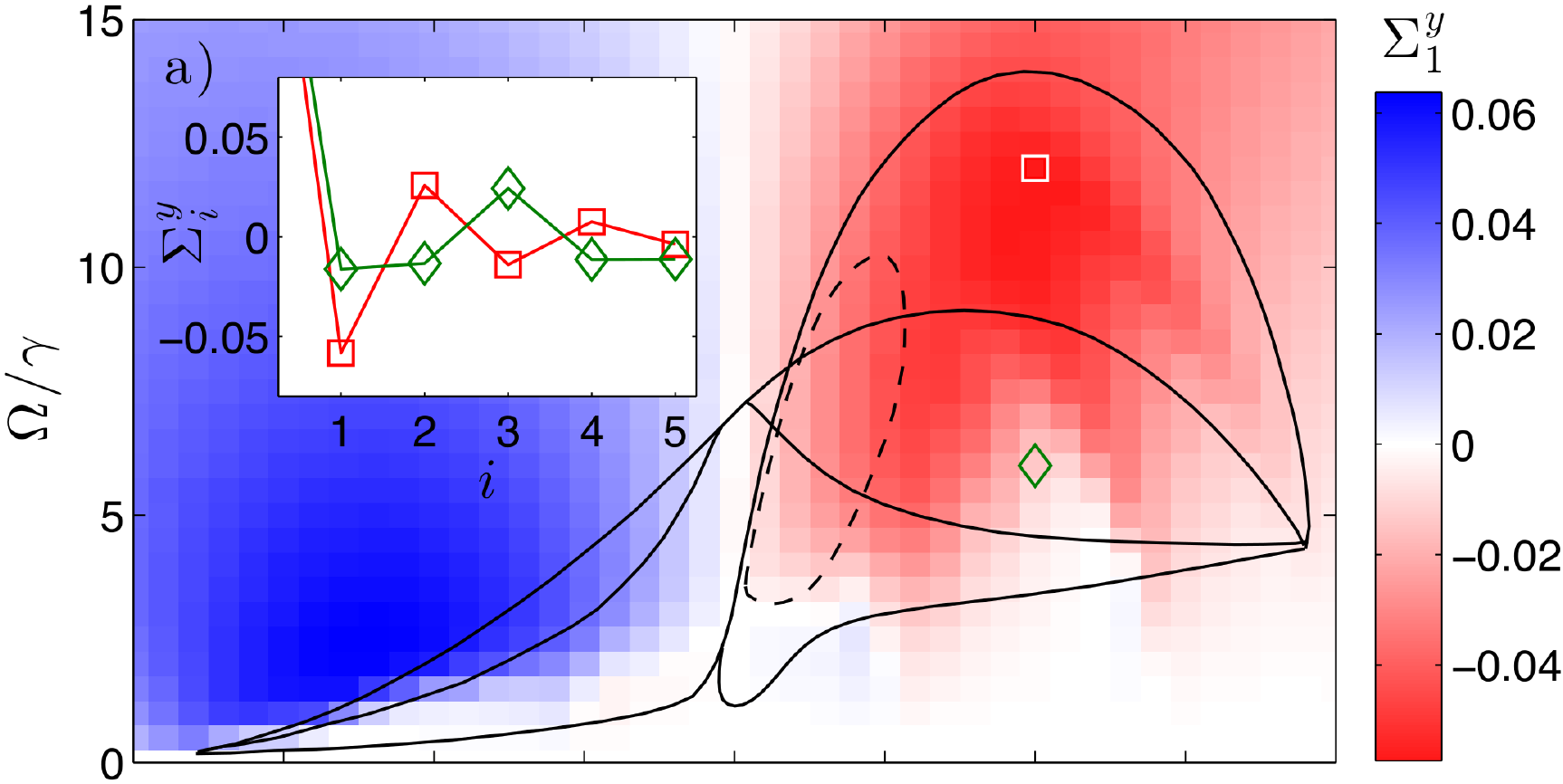} \\  
\includegraphics[width=\columnwidth]{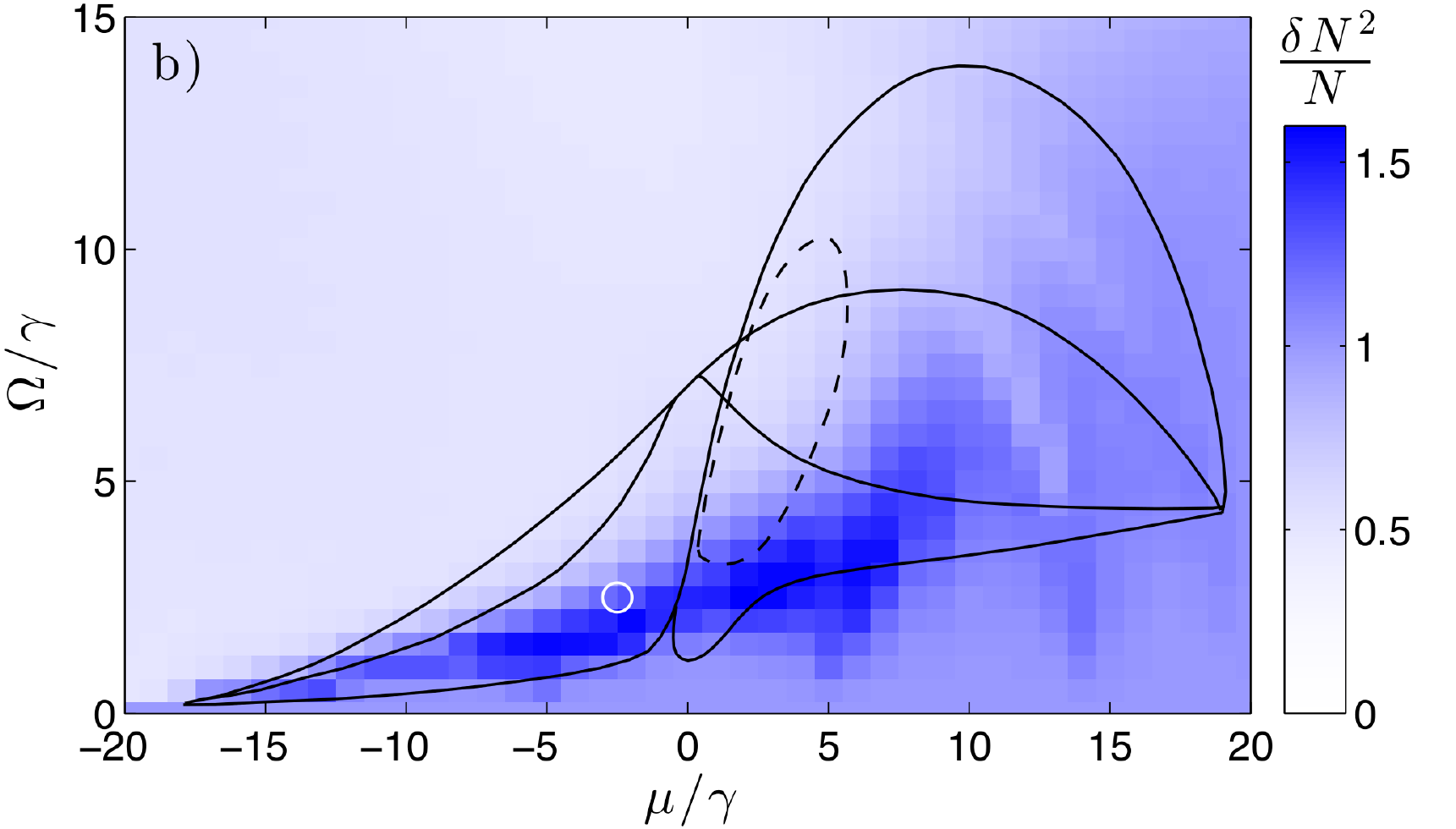} 
\caption{\label{fig2} (color online). Both panels correspond to $\mathcal{N}=12$ cavities with periodic boundary conditions and nearest-neighbor couplings for $J/\gamma = 10$.  The  black lines show the mean-field phase diagram boundaries.  (a)  Nearest-neighbor part of the $\hat{\sigma}^y_i$ correlation function, $\Sigma^y_1$.  The red region indicates antiferromagnetic correlations.   (b) $\delta N^2 / N$ (see text), which is strongly enhanced in the presence of collective bistable switching. The inset in (a) shows $\Sigma^y_i$ for $\mu / \gamma = 10$ and $\Omega / \gamma = 6$ ($12$), exhibiting short-range incommensurate spin density wave (antiferromagnetic) order.}
\end{figure}

We present results for a system with $\mathcal{N}=12$ cavities and periodic boundary conditions in Fig.~\ref{fig2}.  In panel (a), we show the nearest-neighbor part of the $\hat{\sigma}^y$ connected correlation function $\Sigma^y_1$, where $\Sigma^y_i = \langle  \hat{\sigma}^y_j \hat{\sigma}^y_{j+i} \rangle - \langle  \hat{\sigma}^y_j \rangle \langle \hat{\sigma}^y_{j+i} \rangle $, which is independent of $j$.  We choose to study $\hat{\sigma}^y$ correlations because the $y$-components of the spins exhibit the strongest AF order in the MF results, and $\Sigma^y_i$ can be measured via correlated homodyne detection in an experimental setup.  The blue region shows where $\Sigma^y_1$ is positive, and the red region shows where $\Sigma^y_1$ is negative, corresponding to AF nearest-neighbor correlations.   The inset shows $\Sigma^y_i$ for $\mu/\gamma = 10$ and $\Omega / \gamma = 6$ $(12)$, represented by green diamonds (red squares).  While both correlation functions exhibit nearest-neighbor AF order, the correlations for $\Omega / \gamma = 6$ have incommensurate spin density wave character while the correlations for $\Omega / \gamma = 12$ have a  true AF character, switching from negative to positive values with each successive cavity.  The incommensurate spin density wave correlations are present where the MF linear stability analysis predicts an incommensurate spin density wave instability of the $\mathrm{U}_1$ steady state, corresponding to the f-AF region in Fig.~\ref{fig1}.  

We performed finite size scaling calculations for these and a number of other parameters, and find that $\mathcal{N}=12$ accurately captures AF order of the DDBH model in the thermodynamic limit for a large region of the phase diagram; this is due to the short-range, exponentially decaying nature of $\Sigma^y_i$.   The black lines overlaid in Fig.~\ref{fig2} show the MF phase diagram boundaries.    Interestingly, there is reasonably good  agreement between the region of short-range AF correlations in the 1D quantum system and the MF results.

\begin{figure}[t]
\includegraphics[width=\columnwidth]{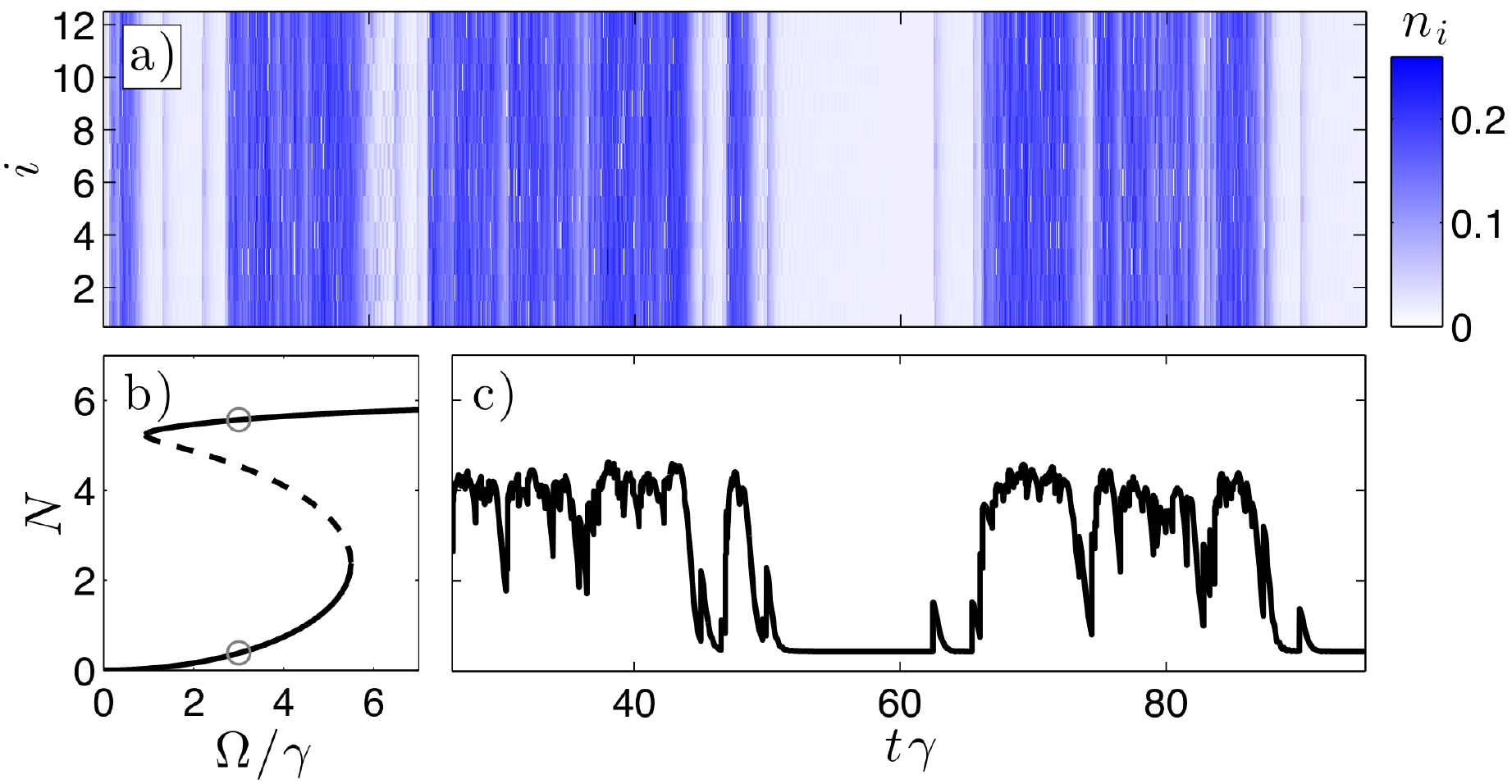}
\caption{\label{fig3} (color online).  All panels correspond to $\mathcal{N}=12$ cavities and $\mu/ \gamma = -2.5$.  (a)  Example quantum trajectory for $\Omega / \gamma = 2.5$, in the bistable regime, showing collective switching between dark (white) and bright (blue) states.  (b)  Mean-field calculation of total photon number, showing collective bistability.  The solid (dashed) black lines correspond to stable (unstable) solutions of the mean-field equations.  (c) Total photon number for the quantum trajectory shown in (a). } 
\end{figure}

In Fig.~\ref{fig2}(b), we show the normalized fluctuations in the total photon number $\hat{N} = \sum_{i=1}^\mathcal{N} \hat{n}_i$, given by $\delta N^2 / N$ where $\delta N^2 = \langle \hat{N}^2 \rangle - \langle \hat{N} \rangle^2 $ and $N = \langle \hat{N} \rangle$.  Interestingly $\delta N^2/N$ becomes anomalously large in the darker blue region, which has significant overlap with the MF bistability, indicating that photon number fluctuations become strongly correlated when the MF theory predicts collective bistability.   The origin of the $\delta N^2 / N$ enhancement is revealed upon inspection of the trajectories themselves.   In Fig.~\ref{fig3}(a), we show the photon number as a function of time for $\mathcal{N}=12$ cavities with $\mu/\gamma = -2.5$ and $\Omega / \gamma = 2.5$, in the region of enhanced $\delta N^2 / N$ (shown by the white circle in Fig.~\ref{fig2}(b)).  Interestingly, the trajectory exhibits collective switching between macroscopically distinguishable states, which  resemble the MF steady states for these parameters.   We plot the total photon number obtained via MF calculations  as a function of $\Omega / \gamma$ for $\mu / \gamma = -2.5$ in panel (b) of Fig.~\ref{fig3}; the values at $\Omega / \gamma = 2.5$ are indicated by the grey markers.  We show $N$ for the quantum trajectory as a function of time in panel (c).  Here, $N$ fluctuates about two mean values for extended periods of time, which are interrupted by switching events that drive the system from one MF-like state to the other.  

\section{Collective bistability \& switching}  A recent work~\cite{Mendoza15} showed that collective bistability in the driven-dissipative XY spin-$\frac{1}{2}$ model vanishes as the Gutzwiller approximation is systematically improved.  Our results demonstrate that while $\hat{\rho}^\mathrm{ss}$ is indeed unique~\cite{Spohn77,Schirmer10}, it retains clear signatures of the MF bistability, namely that the system dynamically switches between two macroscopically distinguishable configurations that resemble the MF steady states $\hat{\rho}^\mathrm{1}$ and $\hat{\rho}^\mathrm{2}$; in fact, this is a generic feature of bistable systems~\cite{Armen06,Kerckhoff11,Lee12,Hu13,Dombi13}.
The approach to $\hat{\rho}^\mathrm{ss}$ in the bistable regime is then characterized by switching between these two states, and the rate of convergence is directly related to the average time spent in $\hat{\rho}^\mathrm{1}$ and $\hat{\rho}^\mathrm{2}$ between switching events; we refer to these times as $\tau_1$ and $\tau_2$, respectively.  In the master equation formalism, the asymptotic rate of convergence is set by the Liouvillian gap $\Gamma = -\mathrm{Re}[\varepsilon_\mathrm{ex}]$, where $\varepsilon_\mathrm{ex}$ is the eigenvalue of ${\mathscr{L}}$ with the smallest magnitude non-zero real part~\cite{Verstraete09,Cai13}; this suggests that $\Gamma$ and $\tau_{1,2}$ are intimately related~\cite{Kinsler91}.   The presence of true bistability in the MF theory reflects the fact that the Liouvillian gap vanishes at this level of approximation.  The collective switching in the exact dynamics, and the concomitant opening of the Liouvillian gap, can be understood to result from both quantum and dissipation-induced (classical) fluctuations that are not included in the MF theory.

\begin{figure}[t]
\includegraphics[width=\columnwidth]{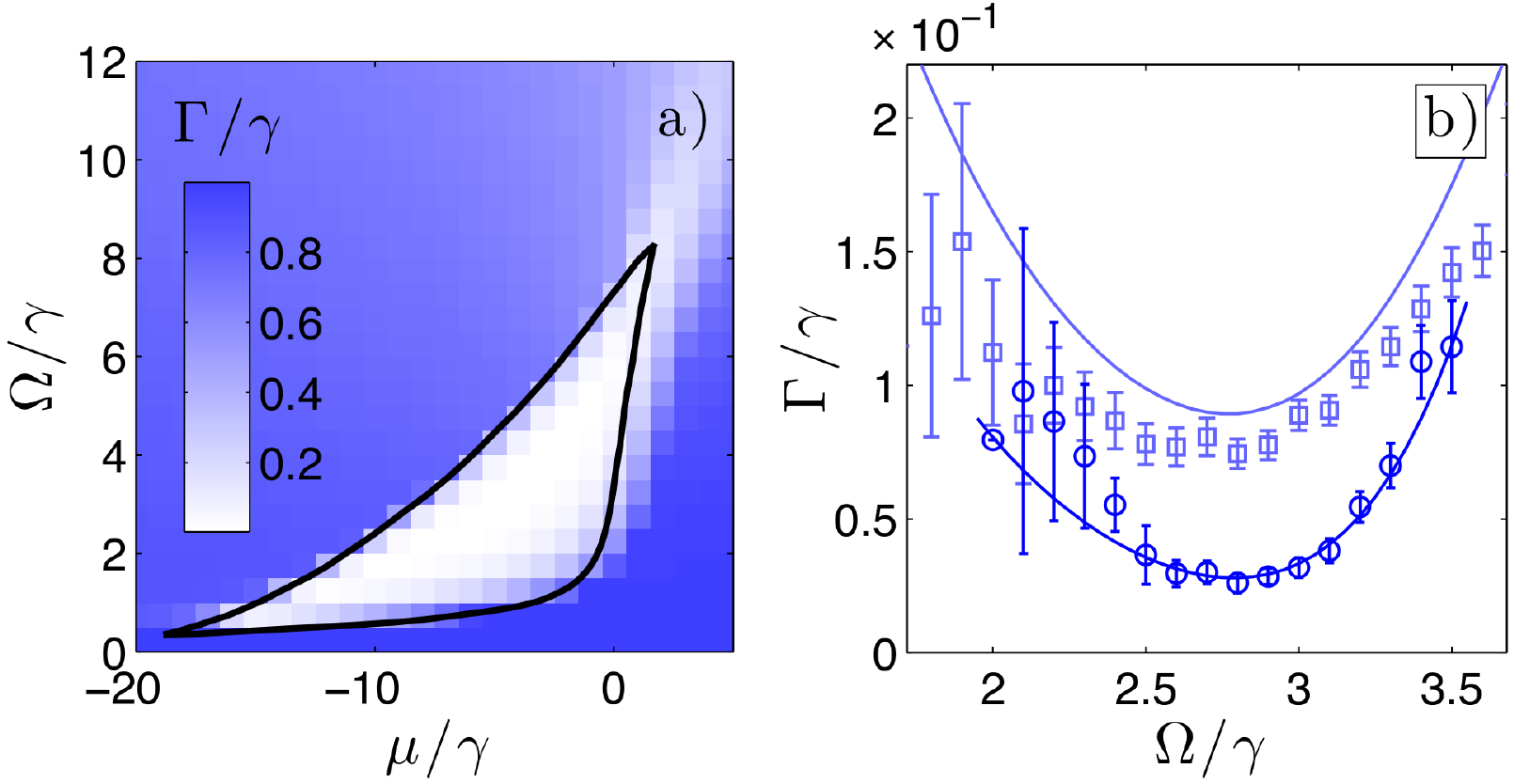} 
 \caption{\label{fig4} (color online). (a) Liouvillian gap $\Gamma / \gamma$ for $\mathcal{N}=20$ cavities with $J/\gamma = 10$ and infinite-range coupling in the hard-core ($U\rightarrow \infty$) limit.  The solid black line shows the mean-field bistable phase boundary for spatially uniform states.  (b) Liouvillian gap for $\mu / \gamma = -5$.  The solid dark (light) blue line shows the gap calculated by diagonalizing the Liouvillian  for $\mathcal{N}=12$ ($8$) cavities.  The circles (squares) show the gap extracted from quantum trajectories simulations (see text); the error bars represent standard error. } 
\end{figure}

It is natural to expect that MF behavior can be recovered in the quantum system as its coordination number $z$ is increased.  We explore this possibility by taking the spin couplings (or photon hopping) to be infinite-range.  This corresponds to modifying the hopping term in Eq.~(\ref{Hamiltonian}) to $- \mathcal{J} \sum_{i \neq j} \hat{a}^\dagger_i \hat{a}_j$, where $\mathcal{J} \equiv 2 J / (\mathcal{N}-1)$.   
While this limit may seem unnatural for arrays of photonic cavities, it could be achieved using an external mirror, and  it is in fact quite natural for other open quantum systems, such as ensembles of Rydberg atoms~\cite{Lee11,Hoening14,Weimer15} or trapped ions~\cite{Porras04,Kim10}.  We calculate the Liouvillian gap exactly by taking advantage of an efficient parameterization of the accessible space of density matrices (see~\cite{Sarkar87,Xu13} and Sec.~\ref{sec:inf} in the appendix); we show the Liouvillian gap for $\mathcal{N}=20$ as a function of $\mu / \gamma$ and $\Omega / \gamma$ in Fig.~\ref{fig4}(a).  There is striking agreement between the exact quantum results and the MF results; where MF theory predicts collective bistability, the Liouvillian gap decreases to $\Gamma \ll \gamma$~\cite{Sarkar87}.

We plot $\Gamma / \gamma$ for $\mathcal{N}=8$ (12) cavities and $\mu / \gamma = -5$ as a function of $\Omega / \gamma$ in Fig.~\ref{fig4}(b), shown by the light (dark) blue solid lines. 
Savage and Carmichael proposed a two-state toy model to describe a bistable system with a small Liouvillian gap~\cite{Savage88}, which has a gap $\Gamma_\mathrm{toy} = \tau_1^{-1}+ \tau_2^{-2}$.    
We extract values for $\tau_1$ and $\tau_2$ heuristically by measuring the time spent in the dark (1) and bright (2) states of the quantum trajectory simulations, and plot $\tau_1^{-1} +  \tau_2^{-1}$ in Fig.~\ref{fig4}(b), shown by the squares (circles) for $\mathcal{N}=8$ ($12$).  Already at $\mathcal{N}=12$, there is excellent quantitative agreement between the exact Liouvillian gap and the results of the simple two state model as extracted from the quantum trajectories with infinite-range couplings.   This provides a clear connection between the MF and quantum solutions in the bistable regime.  We note that Kinsler and Drummond performed a similar analysis for the single-mode quantum parametric oscillator, and also found good quantitative agreement  for large photon numbers~\cite{Kinsler91}.

\section{Discussion}  The thermodynamic limit of the 1D DDBH model with nearest-neighbor interactions is challenging to study numerically, but we expect its dynamics to exhibit collective switching over finite-size domains.  This behavior is reminiscent of equilibrium systems that, while exhibiting a first-order phase transition in higher dimensions, fail to do so in 1D.  Whether or not mean-field bistability is associated with a true first-order phase transition in higher spatial dimensions is an interesting question that warrants further study.  Finally, we note that we have studied the soft-core DDBH model (with finite $U$), and identified features that are directly analogous to those discussed here.  A comprehensive study of the soft-core DDBH model is the subject of future work.

\begin{acknowledgments}
We thank Mohammad Maghrebi, Sarang Gopalakrishnan, and Jeremy Young for insightful discussions. RW, KM, AG, and MH thank the KITP for hospitality.  We acknowledge partial support from the NSF, ONR, ARO, ARL, AFOSR, NSF PIF, NSF PFC at the JQI, and the Sloan Foundation.   This research was supported in part by the NSF under Grant No. NSF PHY11-25915.
\end{acknowledgments}

\appendix

\section{Equations of motion in the $U\rightarrow\infty$ limit}
\label{sechardcore}

In this appendix, we provide technical details to support the theory and numerics in the main text.   Though many of our results are valid more generally for soft-core bosons, here we specialize to the limit of hard-core bosons, valid when $U \rightarrow \infty$, where $U$ is the local interaction strength in Eq.~(2) of the main text.  Hard-core bosons can be conveniently mapped onto spins via an inverse Holstein-Primakoff transformation,
\begin{align}
\label{sHP}
\hat{a}^\dagger_i \rightarrow \hat{\sigma}^+_i~~{\rm and}~~\hat{a}_i \rightarrow \hat{\sigma}^-_i,
\end{align}
where $\hat{\sigma}^\pm_i = (\hat{\sigma}^x_i \pm i \hat{\sigma}^y_i)/2$ and $\hat{\sigma}^+_i \hat{\sigma}^-_i = (\hat{\sigma}^z_i + \mathds{1} )/2$, $\mathds{1}$ is the $SU(2)$ identity operator, and $\hat{\sigma}^{x,y,z}_i$ are Pauli matrices that act on cavity $i$.  Following this transformation, the master equation  becomes
\begin{align}
\label{smaster}
\partial_t \hat{\rho} = -i \big[ \hat{\mathcal{H}} ,\hat{\rho} \big] + \frac{\gamma}{2} \sum_{i} \left( 2 \hat{\sigma}^-_i \hat{\rho} \hat{\sigma}^+_i - \hat{\sigma}^+_i \hat{\sigma}^-_i \hat{\rho} - \hat{\rho} \hat{\sigma}^+_i \hat{\sigma}^-_i \right),
\end{align}
where $\hat{\mathcal{H}}$ is the system Hamiltonian,
\begin{align}
\label{sH}
\hat{\mathcal{H}} = -\frac{J}{ 2 z} \sum_{\langle i , j \rangle} \left( \hat{\sigma}^x_i \hat{\sigma}^x_j + \hat{\sigma}^y_i \hat{\sigma}^y_j   \right)  + \Omega \sum_i \hat{\sigma}^x_i - \frac{\mu}{2} \sum_i \hat{\sigma}^z_i.
\end{align}
Here the summations run over $i=1,\dots,\mathcal{N}$, and the notation $\langle i , j \rangle$ indicates an additional sum over all cavities $j$ that are coupled to cavity $i$; the number of cavities $j$ are quantified by the coordination number $z$.  Equations~(\ref{smaster}) and (\ref{sH}) describe a driven-dissipative spin-$\frac{1}{2}$ XY model, with isotropic interactions and in the presence of a homogeneous applied field.

To study the steady states of Eq.~(\ref{smaster}), it is convenient to use equations of motion for the spin components $\sigma^{x,y,z}_i = \langle \hat{\sigma}^{x,y,z}_i \rangle$; these equations are readily derived by taking $\partial_t \sigma^{x,y,z}_i = \mathrm{Tr} [ \hat{\sigma}^{x,y,z}_i \partial_t \hat{\rho} ]$.  Using the Pauli matrix commutation relations $[ \hat{\sigma}^\alpha_i , \hat{\sigma}^\beta_j ] = 2 i \delta_{i j} \varepsilon_{\alpha \beta \gamma} \hat{\sigma}^\gamma_ i $ where $\varepsilon_{\alpha \beta \gamma}$ is the Levi-Civita symbol, we find
\begin{align}
\label{sspineom}
\partial_t \sigma^x_i &= - \frac{2 J}{z} \sum_{\langle i , j \rangle} \langle \hat{\sigma}^z_i  \hat{\sigma}^y_j \rangle   + \mu \sigma^y_i - \frac{\gamma}{2} \sigma^x_i \nonumber \\
\partial_t \sigma^y_i &=  \frac{2 J}{z} \sum_{\langle i , j \rangle} \langle \hat{\sigma}^z_i  \hat{\sigma}^x_j \rangle - \mu \sigma^x_i - 2 \Omega \sigma^z_i - \frac{\gamma}{2} \sigma^y_i \nonumber \\
\partial_t \sigma^z_i &= \frac{2 J}{z} \sum_{\langle i, j \rangle} \langle   \hat{\sigma}^x_i \hat{\sigma}^y_j - \hat{\sigma}^y_i \hat{\sigma}^x_j  \rangle +  2 \Omega \sigma^y_i -  \gamma (\sigma^z_i + 1).
\end{align}

\subsection{Gutzwiller mean-field approximation}

In the Gutzwiller mean-field (MF) approximation, the density matrix is assumed to factorize over all cavities, $\hat{\rho} = \bigotimes_i \hat{\rho}_i $.  In the spin formalism, this corresponds to the factorization of all non-local two-spin expectation values.  Thus, in the MF approximation, Eqs.~(\ref{sspineom}) become
\begin{align}
\label{sMFspineom}
\partial_t \sigma^x_i &= - \frac{2 J}{z} {\sigma}^z_i  \sum_{\langle i , j \rangle} {\sigma}^y_j    + \mu \sigma^y_i - \frac{\gamma}{2} \sigma^x_i \nonumber \\
\partial_t \sigma^y_i &=  \frac{2 J}{z} {\sigma}^z_i  \sum_{\langle i , j \rangle}  {\sigma}^x_j  - \mu \sigma^x_i - 2 \Omega \sigma^z_i - \frac{\gamma}{2} \sigma^y_i \nonumber \\
\partial_t \sigma^z_i &= \frac{2 J}{z}  {\sigma}^x_i \sum_{\langle i, j \rangle}  {\sigma}^y_j -  \frac{2 J}{z} {\sigma}^y_i \sum_{\langle i, j \rangle} {\sigma}^x_j   +  2 \Omega \sigma^y_i -  \gamma (\sigma^z_i + 1).
\end{align}
In the majority of the phase diagram in Fig.\ 1 of the main text, the steady states are obtained by evolving these equations numerically using a $4^\mathrm{th}$ order Runge-Kutta algorithm.

\begin{figure*}[t]
% \centering
\includegraphics[width=2\columnwidth]{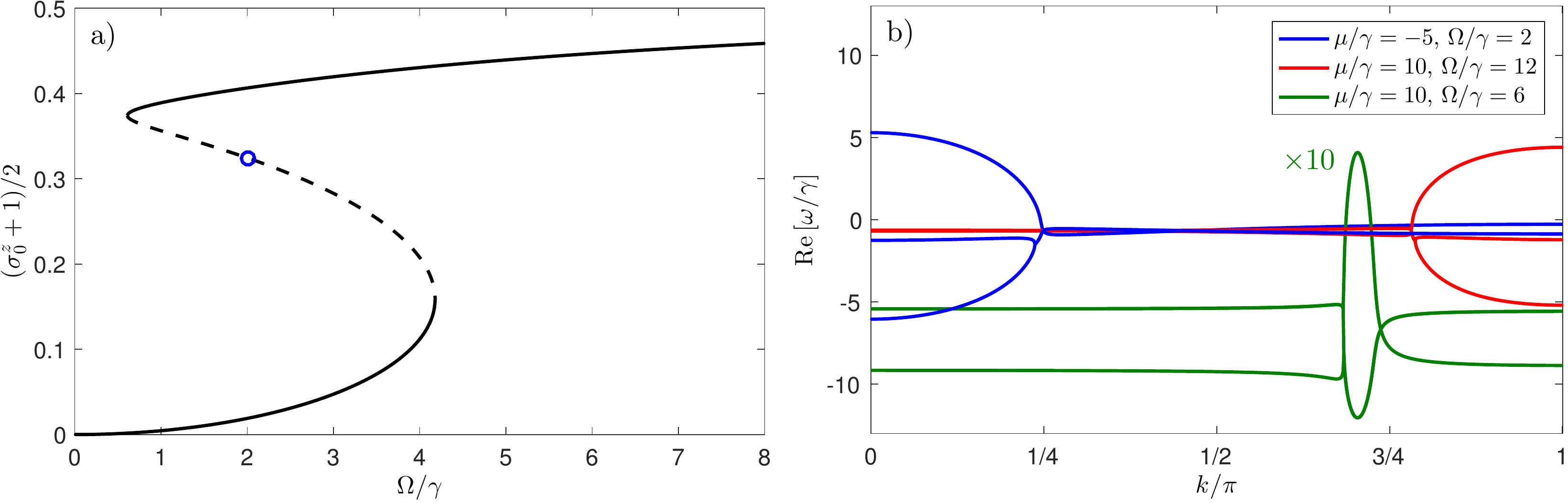}
\caption{\label{suppfig1} a) Gutzwiller mean-field result for $(\sigma^z_0+1)/2$ (equivalent to local photon number) for $J/\gamma = 10$ and $\mu/ \gamma = -5$.  The solid black lines represent steady state solutions that are stable at $k=0$; the dashed black line represents a solution of the mean-field equations that is unstable at $k=0$.  b) Real part of the linear stability eigenvalues as a function of wave number $k$.  The values of $k$ for which $\mathrm{Re}[\omega / \gamma] > 0$ indicates an instability of the uniform steady state  with these values of $k$.   } 
\end{figure*}

\subsection{Linear stability analysis}
\label{sec:linstab}

The phase diagram in Fig.\ 1 of the main text also shows results from a linear stability analysis on spatially uniform steady states within the MF approximation, which is carried out as follows.
Spatially uniform states have the property $\sigma^\alpha_i = \sigma^\alpha_j \equiv \sigma^\alpha$ for all $i,j$.  Using this as an \emph{ansatz} for Eqs.~(\ref{sMFspineom}), we find the following equations for the uniform steady states,
\begin{align}
\label{suniform}
0 &= - 2 J {\sigma}^z  {\sigma}^y + \mu \sigma^y  - \frac{\gamma}{2} \sigma^x \nonumber \\
0 &=  2 J \sigma^z   {\sigma}^x  - \mu \sigma^x - 2 \Omega \sigma^z - \frac{\gamma}{2} \sigma^y \nonumber \\
0 &=  2 \Omega \sigma^y -  \gamma (\sigma^z + 1),
\end{align}
which can be solved analytically to find the spin configurations of the uniform steady states.  For example, we plot solutions for  $J/\gamma = 10$ and $\mu / \gamma = -5$ as a function of $\Omega / \gamma $ in Fig.~\ref{suppfig1}(a); these solutions exhibit collective bistability, where the dynamically stable steady state solutions are shown by the solid black lines. The dashed black line in this figure indicates a dynamically unstable solution, which we discuss in more detail below.

The \emph{ansatz} of spatially uniform steady states is not appropriate under all circumstances, for example, if the steady state develops a spatial order that breaks the discrete translational symmetry of the DDBH model.  The presence of such spatially ordered states can sometimes be captured by an instability of the uniform steady state.  To explore this possibility, we perform a linear stability analysis on the uniform steady state solutions of Eqs.~(\ref{suniform}), specializing to one-dimensional systems with $z=2$.  In a system with infinite spatial extent, this is accomplished by adding a small plane wave perturbation to the uniform steady state of the form $\boldsymbol{\sigma}_m = \boldsymbol{\sigma}_0 + \boldsymbol{\delta} e^{ikm}$, where $\boldsymbol{\sigma}_0 = (\sigma^x_0,\sigma^y_0,\sigma^z_0)^\mathrm{T}$ and $\sigma^\alpha_0$ are the solutions of Eqs.~(\ref{suniform}), $m$ are cavity indices, and $k$ is the wave number of the perturbation.  Linearizing in $\boldsymbol{\delta}$, we find the equation
\begin{align}
\label{linstab}
\partial_t \boldsymbol{\delta} = \mathds{M} \boldsymbol{\delta},
\end{align}
where
\begin{widetext}
\begin{align}
\label{M}
\mathds{M} = \left( \begin{array}{ccc} -\frac{\gamma}{2} & \mu - 2 J \sigma^z_0 \cos (k)  & - 2 J \sigma^y_0  \\
 2 J \sigma^z_0 \cos (k) - \mu  & - \frac{\gamma}{2} &  2 J \sigma^x_0 - 2\Omega  \\
 2 J \sigma^y_0 (1 - \cos (k) )   & - 2 J \sigma^x_0 (1 - \cos (k) ) + 2 \Omega   &  -\gamma  \end{array}   \right).
\end{align}
\end{widetext}
The matrix $\mathds{M}$ has eigenvalues $\omega$ that depend on the wave number $k$. When the real part of $\omega$ is negative for all $k$, the uniform steady state $\boldsymbol{\sigma}_0$ is dynamically stable.  When some $\omega$ acquires a positive real part, this signifies an instability of the uniform steady state, and the wave number at which $\mathrm{Re}[\omega]$ is maximum corresponds to the mode that is maximally unstable, and dominates the dynamics of the instability.  
For example, in Fig.~\ref{suppfig1}(b) we plot $\mathrm{Re} [ \omega / \gamma ] $ as a function of $k$ for three sets of parameters, all with $J / \gamma = 10$.  The blue lines correspond to one of three solutions with $\Omega / \gamma = 2$, shown by the blue circle in panel (b) of this figure.  These modes are unstable at $k=0$, indicating a global instability of the uniform steady state solution.  The green and red lines correspond to parameters exhibiting spin density wave instabilities.  The green line exhibits an incommensurate spin density wave instability, as the modes are stable at $k=\pi$ but unstable in a small region of $k<\pi$.  The red line exhibits an antiferromagnetic (AF) spin density instability, as the most unstable eigenvalue occurs for $k=\pi$.  

and antiferromagnetic (AF) instabilities, respectively.  

\section{Exact numerical solution of Eq.~(\ref{smaster})}
\label{sec:jumps}

 We employ a quantum trajectories algorithm to solve the master equation~(\ref{smaster}), which is a powerful, exact method for studying open quantum systems that relies on treating the system-environment coupling (the dissipator in the master equation) stochastically.  Here, we present the algorithm  for generating a quantum trajectory using quantum jumps. We begin by defining the effective non-Hermetian Hamiltonian,
\begin{align}
\hat{\mathcal{H}}_\mathrm{eff} = \hat{\mathcal{H}} - i \frac{\gamma}{2} \sum_i \hat{n}_i ,
\end{align}
so Eq.~(\ref{smaster}) can be written as
\begin{align}
\label{smastereff}
\partial_t \hat{\rho} = -i \left( \hat{\mathcal{H}}_\mathrm{eff} \hat{\rho} - \hat{\rho} \hat{\mathcal{H}}_\mathrm{eff}^\dagger  \right) + \gamma \sum_i \hat{a}_i \hat{\rho} \hat{a}_i^\dagger .
\end{align}
The quantum trajectories method amounts to evolving a wave function, as opposed to a density matrix, under $\hat{\mathcal{H}}_\mathrm{eff}$, while treating the right-most ``recycling'' term in Eq.~(\ref{smastereff}) stochastically.  

Consider a wave function $| \psi (t) \rangle$ at time $t$ that evolves under $\hat{\mathcal{H}}_\mathrm{eff}$.  For a small time step $\delta t$, the wave function at $t+\delta t$ can be written as (using Euler integration)
\begin{align}
\label{sEuler}
| \psi' (t+dt ) \rangle = ( 1- i \hat{\mathcal{H}}_\mathrm{eff} dt ) | \psi(t) \rangle .
\end{align}
The norm of $|\psi \rangle$ is not conserved in real-time evolution under $\hat{\mathcal{H}}_\mathrm{eff}$ due to its being non-Hermetian, so in general we have
\begin{align}
\label{snorm}
\frac{ \langle  \psi^\prime (t+dt)   | \psi^\prime(t+dt)    \rangle }{  \langle \psi(t) | \psi(t) \rangle} = 1 - \delta p .
\end{align}
In the quantum trajectories formalism, we choose the wave function at $t+\delta t$ stochastically.  With probability $1-\delta p$, we choose
\begin{align}
\label{s1}
| \psi (t+\delta t) \rangle = \frac{| \psi^\prime (t+\delta t) \rangle}{\sqrt{1- \delta p}},
\end{align}
and with probability $\delta p$ we take our next state to be one that emitted a photon from cavity $i$,
\begin{align}
\label{s2}
|\psi(t+\delta t) \rangle = \frac{\hat{a}_i | \psi(t) \rangle}{ \sqrt{\delta p_i / \delta t} },
\end{align}
where the jump-site $i$ is chosen with probability
\begin{align}
\Pi_i = \frac{\delta p_i}{\delta p} = \frac{\delta t \langle \psi(t) | \hat{n}_i | \psi(t) \rangle }{\delta p} ,
\end{align}
and $\sum_i \delta p_i = \delta p$.

\section{Infinite range interactions}
\label{sec:inf}

In general, long-range interactions introduce considerable difficulties when numerically studying the dynamics of many-body systems.  However, when all pairs of spins interact with the same strength, the model becomes symmetric under the exchange of any two spins, which allows for an extremely efficient parametrization of the accessible Hilbert space.  In the absence of spontaneous emission, permutation symmetry of the Hamiltonian restricts the dynamics of initially permutation symmetric pure states to the subspace of collective spin-states, often referred to as Dicke states, and the dimension of the accessible Hilbert space is therefore $\mathcal{O} (\mathcal{N})$.  For an open quantum system governed by a permutation symmetric Liouvillian, dynamics is restricted to the space of permutation symmetric density matrices, which can be parametrized in terms of symmetrized direct products of Pauli matrices.  The space of permutation symmetric, Hermitian matrices over the product Liouville space of $\mathcal{N}$ spins is spanned by basis states of the following form,
\begin{widetext}
\begin{align}
\hat{\mathcal{M}}(\mathbf{n}) &= \frac{1}{2^\mathcal{N}} \sum_\chi \left( \hat{\sigma}^x_{\chi(1) } \otimes \cdots \otimes \hat{\sigma}^x_{\chi (n_x) } \right)  \otimes \left(  \hat{\sigma}^y_{\chi(n_x + 1) } \otimes \cdots \otimes \hat{\sigma}^y_{\chi (n_x+n_y) }  \right) \nonumber \\
 &\otimes \left(  \hat{\sigma}^z_{\chi(n_x + n_y + 1) } \otimes \cdots \otimes \hat{\sigma}^z_{\chi (n_x+n_y + n_z) } \right)  \otimes \left(  \hat{\sigma}^0_{\chi(n_x + n_y + n_z + 1) } \otimes \cdots \otimes \hat{\sigma}^0_{\chi (\mathcal{N}) }  \right).
\end{align}
\end{widetext}
Here the parameters $\mathbf{n} = \{ n_x , n_y , n_z \}$ are positive integers that are arbitrary up to the constraint $n_x + n_y + n_z \leq \mathcal{N}$, $\hat{\sigma}^0_i$ is the identity matrix on the Hilbert space of spin $i$, and $\chi$ is a permutation of integers $\{ 1 , \ldots , \mathcal{N} \}$.  The most general permutation symmetric Hermitian matrix can be written 
\begin{align}
\label{srho}
\hat{\rho} = \sum_\mathbf{n} c(\mathbf{n}) \hat{\mathcal{M}}(\mathbf{n}). 
\end{align}
Note that $\hat{\mathcal{M}}(\{ 0, 0 , 0 \})$ has unity trace, while all the other basis states are traceless; hence $\hat{\rho}$ can only be a valid density matrix if $c(\{ 0, 0, 0 \} ) = 1$.  The number of unconstrained coefficients, and hence the dimensionality of the space that must be considered in the dynamics, is the number of ways of choosing $\mathbf{n}$ such that $0 < n_x, n_y,n_z \leq \mathcal{N}$.  It is straightforward to check that there are $\mathcal{D}$ such $\mathbf{n}$, where
\begin{align}
\mathcal{D} = \frac{(\mathcal{N}+3)(\mathcal{N}+2)(\mathcal{N}+1)}{3 !} - 1,
\end{align}
hence the dimension of the accessible space of density matrices is $\mathcal{O}(\mathcal{N}^3)$.  Even after the restriction $c(\{ 0,0,0\})=1$, a matrix of the form in Eq.~(\ref{srho}) is not guaranteed to be a valid density matrix, as it may have negative eigenvalues (equivalently, it need not satisfy $\mathrm{Tr}[\hat{\rho}^2] \leq \mathrm{Tr}[\hat{\rho}] = 1$).  However, the master equation provides a positive map, and preserves the positivity of the density matrix eigenvalues dynamically.  Therefore if we start with an initial state that is a valid density matrix, it will remain restricted to the space of valid density matrices during the time evolution.

The permutation symmetry of the Liouvillian endows it with a block diagonal structure, with one of the blocks spanned by the states $\hat{\rho}$ in Eq.~(\ref{srho}).  Dynamics of an initial density matrix within this block can therefore be calculated by determining the action on all states $\hat{\mathcal{M}}(\mathbf{n})$:
\begin{align}
\label{eq:Ldef}
\mathscr{L}(\hat{\mathcal{M}}(\mathbf{n})) = \sum_\mathbf{m} L_{\mathbf{m},\mathbf{n}} \hat{\mathcal{M}}(\mathbf{m}).
\end{align}
The coefficients $L_{\mathbf{m},\mathbf{n}}$ are straightforward to compute, and using this Eq.\ (\ref{eq:Ldef}) the master equation can be written
\begin{align}
\partial_t c(\mathbf{n}) = L_{\mathbf{n},\mathbf{m} } c(\mathbf{m}).
\end{align}
This set of $\mathcal{O}(\mathcal{N}^3)$ first-order, linear, ordinary differential equations is straightforward to integrate for $\mathcal{N} \lesssim100$.

\end{document}